\documentclass[a4paper,11pt]{article}
\pdfoutput=1 % if your are submitting a pdflatex (i.e. if you have
\usepackage{jinstpub} % for details on the use of the package, please
\title{\boldmath Tracking in dense environments and its inefficiency measurement using pixel $dE/dx$}

\author{Jason D. Mansour}
\affiliation{Institute of High Energy Physics, Chinese Academy of Sciences, Beijing, China}

\emailAdd{jmansour@cern.ch}

\abstract{We present a measurement of the charged particle reconstruction inefficiency inside of jet cores, using data collected by the ATLAS experiment in 2015 of $pp$ collisions produced at the LHC, at a center-of-mass energy of 13$\;$TeV.  The determination of this inefficiency is important for jet energy scale and mass calibration, as well as multiple other performance studies and analyses.  A data driven method is used, where the fraction of lost particle tracks is determined from energy deposition $dE/dx$ in the pixel detector.  The fraction of lost tracks is found to be less than $5\%$, which is an improvement since the previous study, and agrees well within systematic uncertainties with a Monte Carlo simulation.}


\collaboration[c]{on behalf of the ATLAS collaboration}

\proceeding{8$^{\text{th}}$ International Workshop on Semiconductor Pixel Detectors for Particles and Imaging\\
  5--9 September 2016\\
  Sestri Levante, Italy}

\begin{document}
\maketitle

\section{Introduction}
\label{sec:intro}

The performance of charged particle track reconstruction in dense environments, such as the core of high $p_{T}$ jets, is important for many analyses and performance studies.  Examples include $b$-tagging and boosted $\tau$ reconstruction, jet energy scale and mass determination, and analyses using jet substructure information.  The leading source of systematic uncertainty is in many cases the uncertainty on track reconstruction efficiency.  Thus, it is necessary to maintain a high tracking efficiency also inside of jets, and to determine the (in-)efficiency precisely.

Tracks are reconstructed from hit clusters in the inner detector.  Clusters shared between multiple tracks are penalized during reconstruction, to ensure high quality tracks, as one of the tracks in this case is likely to be fake.  However, in dense environments this is a disadvantage, since clusters from close-by tracks can naturally merge.  To account for this, an artificial neural network is trained to identify and not penalize such merged clusters~\cite{tide,tidenn}.  This greatly increases the percentage of correct associations of clusters to tracks at small track separations, and improves $b$-tagging and $\tau$ reconstruction performance.  To determine the remaining inefficiency of tracking in dense environments, a data driven method is applied, where the fraction of lost tracks is determined from the energy deposition $dE/dx$ in the pixel detector (section~\ref{sec:method}).

\newcommand{\coordinates}{ATLAS uses a right handed coordinate system, centered around the nominal interaction point (IP) at the center of the detector.  The $x$ axis points towards the center of the LHC, the $y$ axis points upwards, and the $z$ axis in parallel to the proton beams.  In the transverse plane, cylindrical coordinates $(r,\varphi)$ are used, where $\varphi$ is the polar angle around the $z$ axis.  Instead of azimuthal angle $\theta$, pseudorapidity $\eta = - \ln \tan \theta/2$ is often used.}

The ATLAS~\cite{atlas} pixel detector is part of the inner detector system, together with the semiconductor tracker (SCT) and the transition radiation tracker (TRT).  The pixel detector is built in a barrel and disc geometry, and has a pseudorapidity coverage\footnote{\coordinates} of $|\eta| < 2.5$.  It is built mostly from planar silicon pixel modules.  The insertable B-Layer (IBL), which was added after a long shutdown (2013--2015), includes planar and 3D sensors~\cite{atlasibl}.  There are four barrels, including the IBL, situated at $r=33.2$, $50.5$, $88.5$ and $122.5\;$mm.  The forward region is instrumented with three disk pairs at $z = \pm 495$, $580$ and $650\;$mm.  Sensor pixels are typically $50\;$\textmu{}m in transverse direction and  $400\;$\textmu{}m in longitudinal direction, whereas pixels of the IBL are only $250\;$\textmu{}m longitudinally.

A measure of energy deposition $dE/dx$ is given by the pixel detector time-over-threshold (ToT).  This is the time that a pulse, caused by a particle, spends over a given threshold, and is approximately proportional to the collected charge.  In particle reconstruction, pixels are grouped by a clustering algorithm into clusters.  The $dE/dx$ value of a cluster is determined by the total collected charge.

Since the magnetic field of the detector bends particle trajectories apart as they move out of the detector, nearby clusters are more likely to merge closer to the interaction point.  However, the IBL only encodes ToT information in 4 bits, whereas the next layer, the B-layer, uses 8 bits and provides a better ToT resolution.  For that reason, in the following the $dE/dx$ information from clusters in the B-Layer will be used.

\section{Samples}
\label{sec:samples}
For this analysis, data samples recorded by the ATLAS detector in 2015 (Run II) of proton-proton collisions produced by the LHC at $\sqrt{s}=13\;$TeV were used, corresponding to an integrated luminosity of $2.8\;\mathrm{fb}^{-1}$.  Events were selected passing single jet triggers, with a minimal jet $p_{T}$ threshold of 100\;GeV.  The triggers were subject to a prescaling depending on the instantaneous luminosity and the energy of the jet triggered on.  This suppresses low $p_{T}$ jets, while keeping all events including a jet with at least $p_{T} > 1\;$TeV, leading to a more uniform transverse momentum spectrum.  Events were required to pass standard data quality requirements, and contain at least one reconstructed primary vertex, associated to at least three tracks.

Data is compared with a Monte Carlo simulation, generated by \textsc{Pythia}~8.186~\cite{pythia}.  Generator parameters were set according to the A14 tune for parton showering and hadronization, and parton distribution functions (PDF) were taken from \textsc{NNPDF23LO}~\cite{nnpdf}.  For comparison, samples were also generated using \textsc{Herwig++}~2.7.1~\cite{herwigpp} with the UEEE5 tune and the \textsc{CTEQ6L1} PDF set~\cite{cteq}, as well as \textsc{Sherpa}~2.1~\cite{sherpa} using \textsc{CT10} PDFs~\cite{ct10}.  Events are digitized using a \textsc{GEANT4} based simulation of the ATLAS detector, and then reconstructed using the same reconstruction algorithms as used for data.  Monte Carlo events are finally reweighted to match the number of events triggered on in data.

\section{Object Selection}

Jets used were seeded from topological clusters~\cite{topocell} and reconstructed by the anti-$k_T$ algorithm~\cite{antikt} with a cone radius of $R=0.4$. They were required to have a transverse momentum of $p_T^\mathrm{jet} \geq 200\;\mathrm{GeV}$ and lie in the region of $|\eta^\mathrm{jet}| < 2.5$.  Jets have been calibrated to the hadronic jet energy scale using a calibration derived from Monte Carlo~\cite{jes}.  It has been shown previously that simulated jet properties agree well with data \cite{jetprop}.

Tracks are reconstructed using an iterative algorithm.  They are seeded using combined measurements from the silicon detectors, and reconstructed using a combinatorial Kalman filter together with a stringent ambiguity solver~\cite{kalman,kalmanapplication}.  The following cuts are applied to tracks:

\pagebreak
\begin{itemize}
	\item $p_T^\mathrm{trk} > 10\;\mathrm{GeV}$
	\item $|\eta^\mathrm{trk}| < 1.2$
	\item $|d_0^\mathrm{BL}| < 1.5\;\mathrm{mm}\,,$ where $d_0^\mathrm{BL}$ is the transverse impact parameter w.r.t. the beamline position
	\item $|z_0^\mathrm{BL} \sin \theta| < 1.5\;\mathrm{mm}\,,$ where $z_0^\mathrm{BL}$ is the distance in $z$ direction between the track's point of closest approach and the primary vertex, and $\theta$ is the polar angle of the track at this point
	\item Number of SCT hits $\geq 6$
	\item Number of pixel holes\footnote{A pixel hole is defined as a expected hit, where the reconstructed track crosses the detector surface, but no hit is recorded.  Inactive parts such as sensor edges or disabled modules are excluded from the definition and do not create holes.} $\leq 1$
\end{itemize}

\section{Template Fit Method}
\label{sec:method}
The goal of this method is to determine the fraction of tracks lost due to merged clusters in jet cores.  The energy deposition $dE/dx$ of pixel clusters follows a Landau distribution~\cite{PDG}, assuming the material is sufficiently thin and only single particles hit the clusters.  The peak of the distribution is around the minimally ionizing particle (MIP) energy.  In the case where two particles contribute to the same cluster, a second peak at $2\times$ the MIP energy is visible.  A third weaker peak can appear for three particles hitting the same cluster.

When a cluster is assigned to only one reconstructed track (not multiply used), two situations are possible: The cluster was indeed hit by only one particle, or it was hit by one reconstructed particle, and another missed one.  It is impossible to distinguish both situations on a per-cluster basis, but one can determine statistically how often each situation occurs, by comparing the two peaks in the $dE/dx$ distribution.  From this, the probability that a track is lost due to merging can be computed.

\begin{figure}
\centering
% \fbox
{\includegraphics[width=0.6\textwidth,trim=0 0 0 1.5cm]{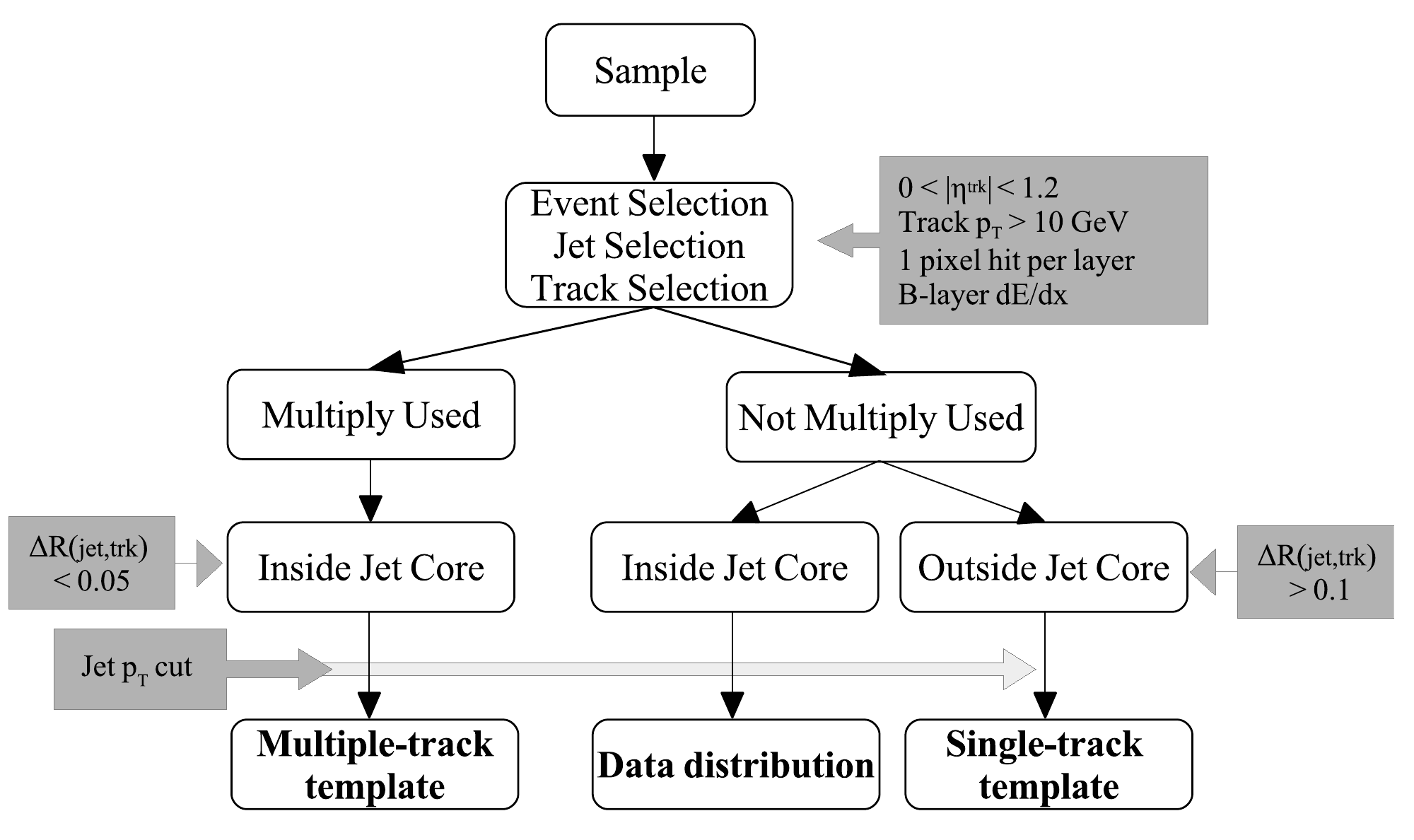}}
\caption{\label{fig:def} Definition of template and data distributions.  This and all further figures taken from~\cite{pubnote}}
\end{figure}

\begin{figure}
\centering
% \fbox{}
\includegraphics[width=0.45\textwidth]{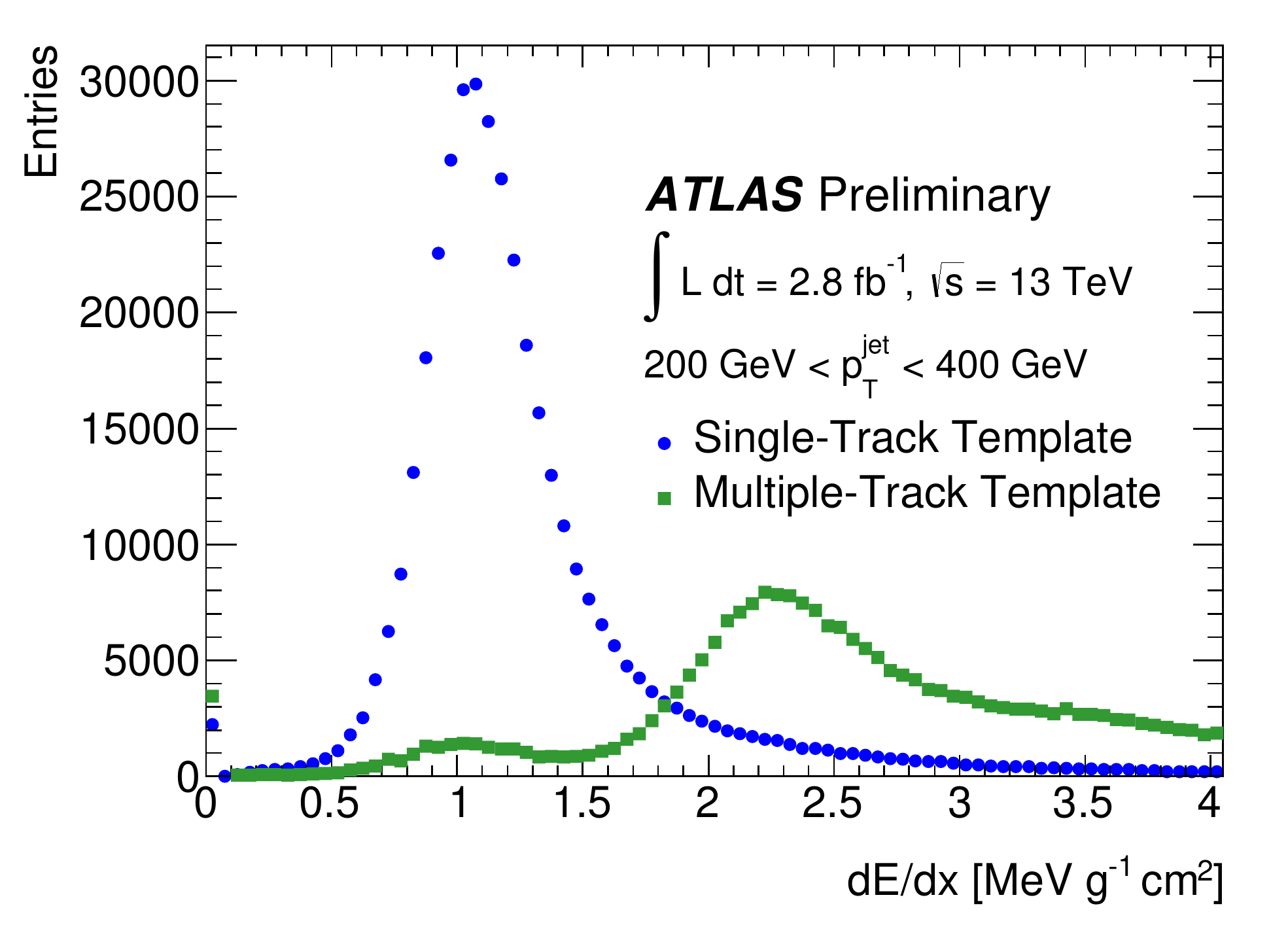}
\includegraphics[width=0.45\textwidth]{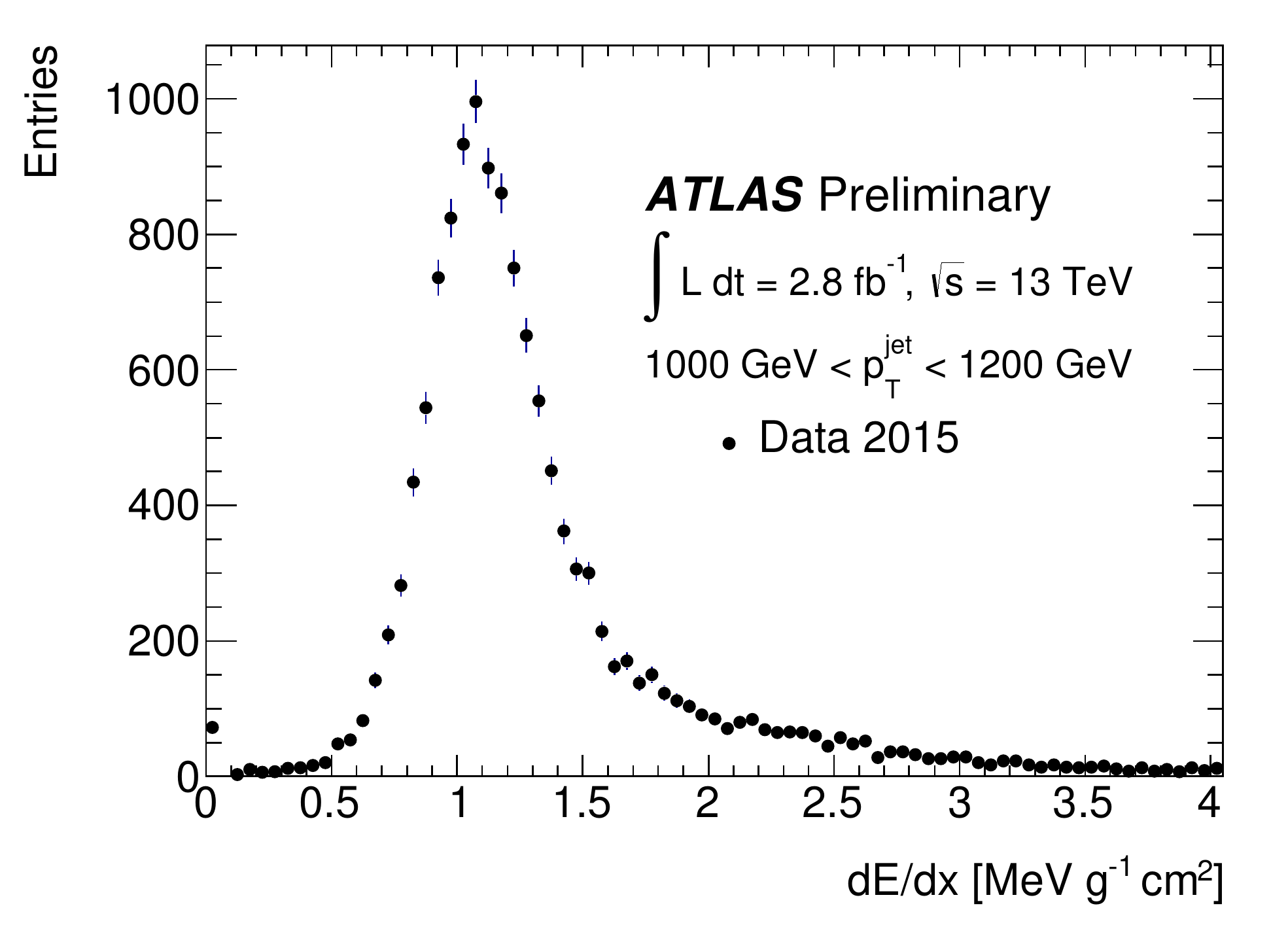}
\caption{
\label{fig:templates}
Left: Single and multiple track templates, derived from data.  Right: Energy loss $dE/dx$ distribution from pixel clusters in jet cores, to be fit with the templates.  From~\cite{pubnote}.}
\end{figure}

To determine the individual contributions of both cases, a template fit is used.  The data to be fitted is the $dE/dx$ distribution of not multiply used clusters, in the core of jets (angular separation\footnote{Angular distance is given by $\Delta R = \sqrt{\Delta \varphi^2 + \Delta \eta^2}$} between center of jet and track $\Delta R(\mathrm{trk}, \mathrm{jet}) < 0.05$).  To get a sample that is enriched in clusters hit by single tracks (single-track template), a selection outside of the jet core ($\Delta R(\mathrm{trk}, \mathrm{jet}) > 0.1$) is applied.  An enriched multiple-track template is obtained by staying in the jet core, but using multiply-used clusters instead (that is, clusters that have multiple confirmed reconstructed tracks).  The selection is summarized in figure~\ref{fig:def}.  Data is separated into seven $p_{T}^\mathrm{jet}$ bins ranging from $200$--$1600\;$GeV.  Since at high $p_T^\mathrm{jet}$ the available statistics is low, the templates taken at $p_T^\mathrm{jet}=200$--$400\;$GeV are used to fit all distributions in data.  The single- and multiple-track templates, and a data distribution can be seen in figure~\ref{fig:templates}.  To minimize the influence of clusters which were hit by three tracks, the fit was performed in a reduced region of $0.8$--$3.2 \;\mathrm{MeV}\,\mathrm{g}^{-1}\,\mathrm{cm}^{2}$ for MC, and $0.67$--$3.07 \;\mathrm{MeV}\,\mathrm{g}^{-1}\,\mathrm{cm}^{2}$ for data.  The regions were chosen such that the fraction of all clusters they contain is the same in data as in MC.

\begin{figure}
\centering
\includegraphics[width=0.45\textwidth]{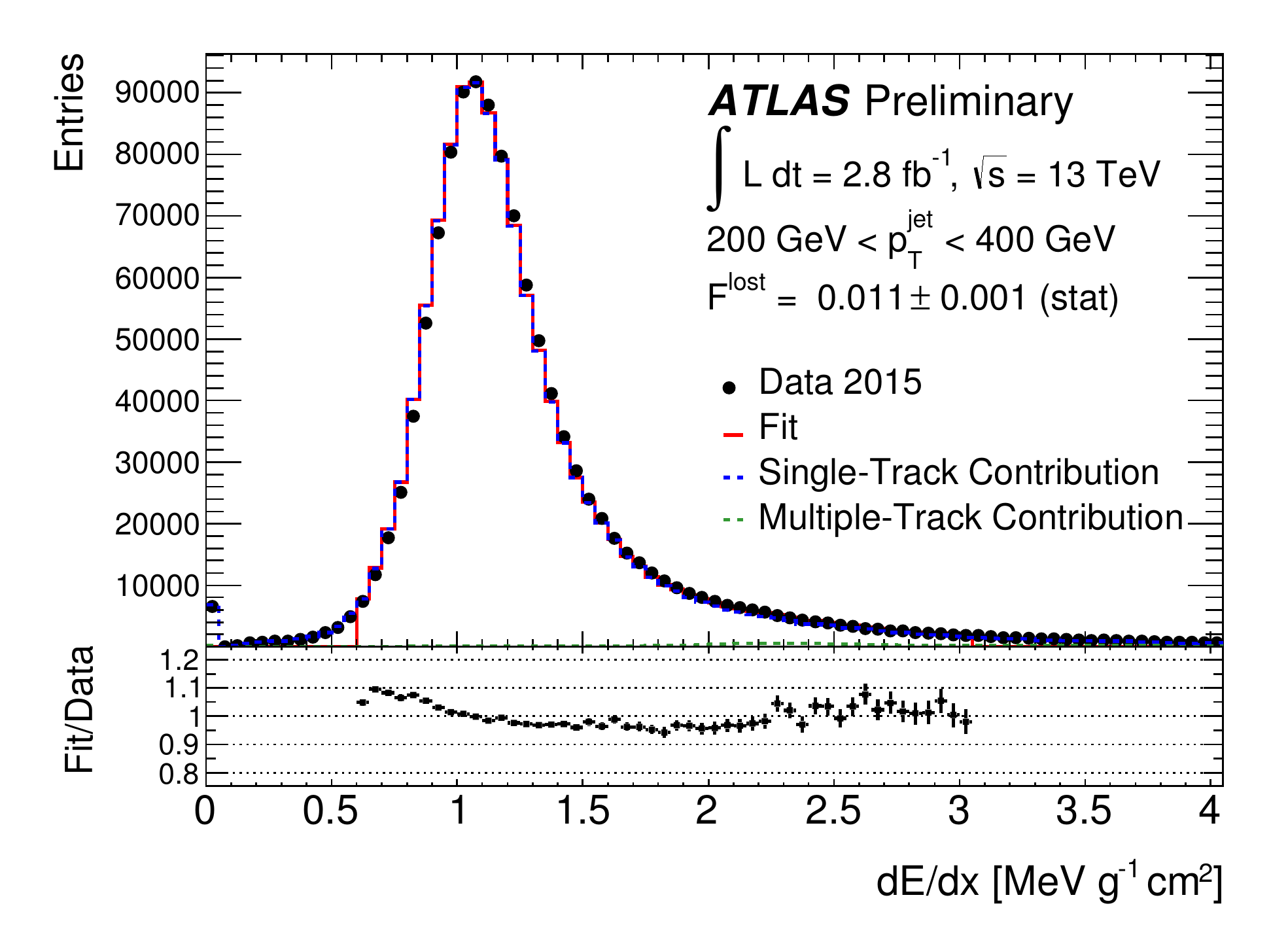}
\includegraphics[width=0.45\textwidth]{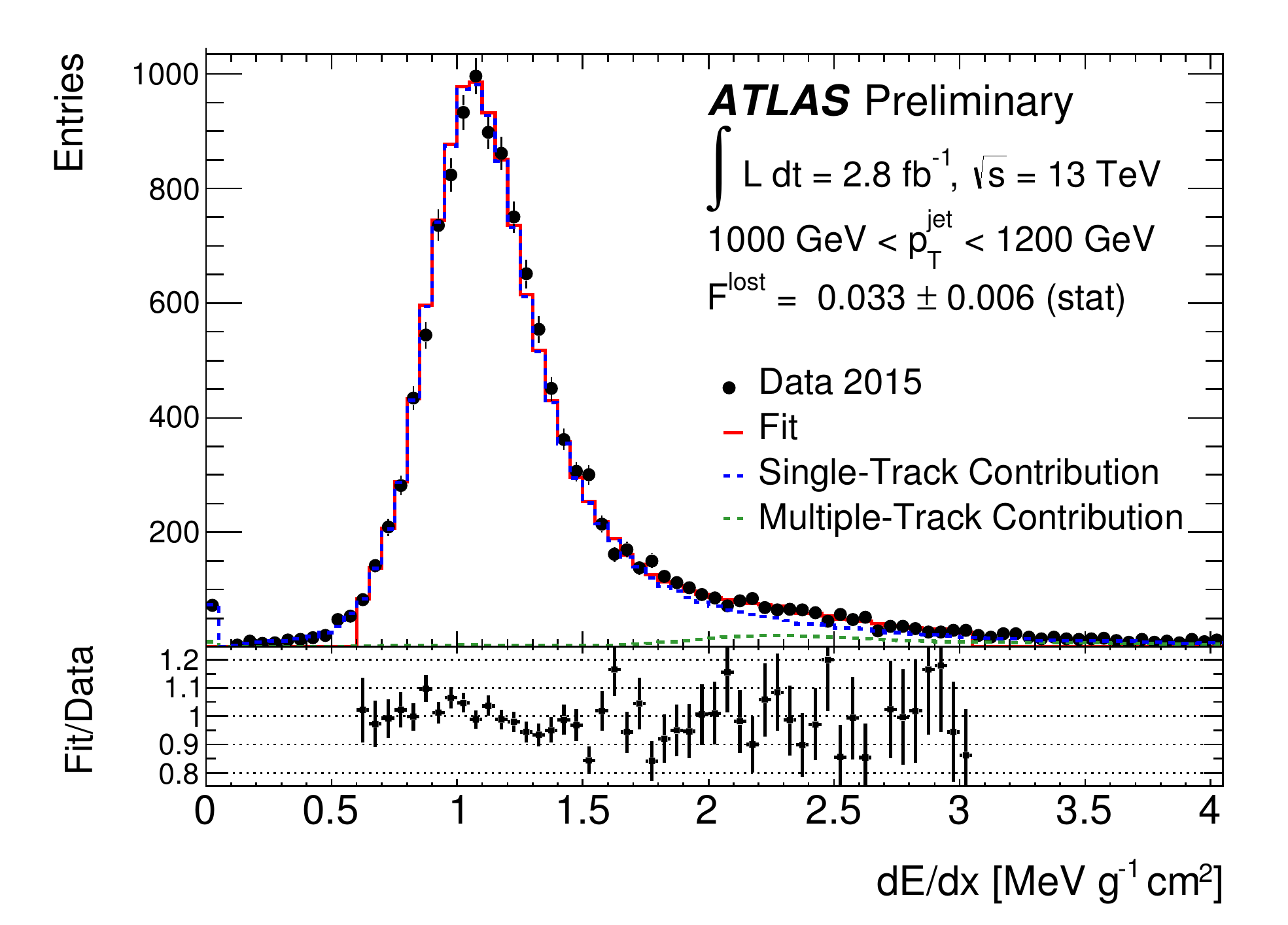}
\caption{\label{fig:datafit}
Result of the template fit for data, for values of jet $p_T=200$--$400\;$GeV (left) and $p_T=1000$--$1200\;$GeV (right).  From~\cite{pubnote}.}
\end{figure}

\begin{figure}
\centering
\includegraphics[width=0.7\textwidth]{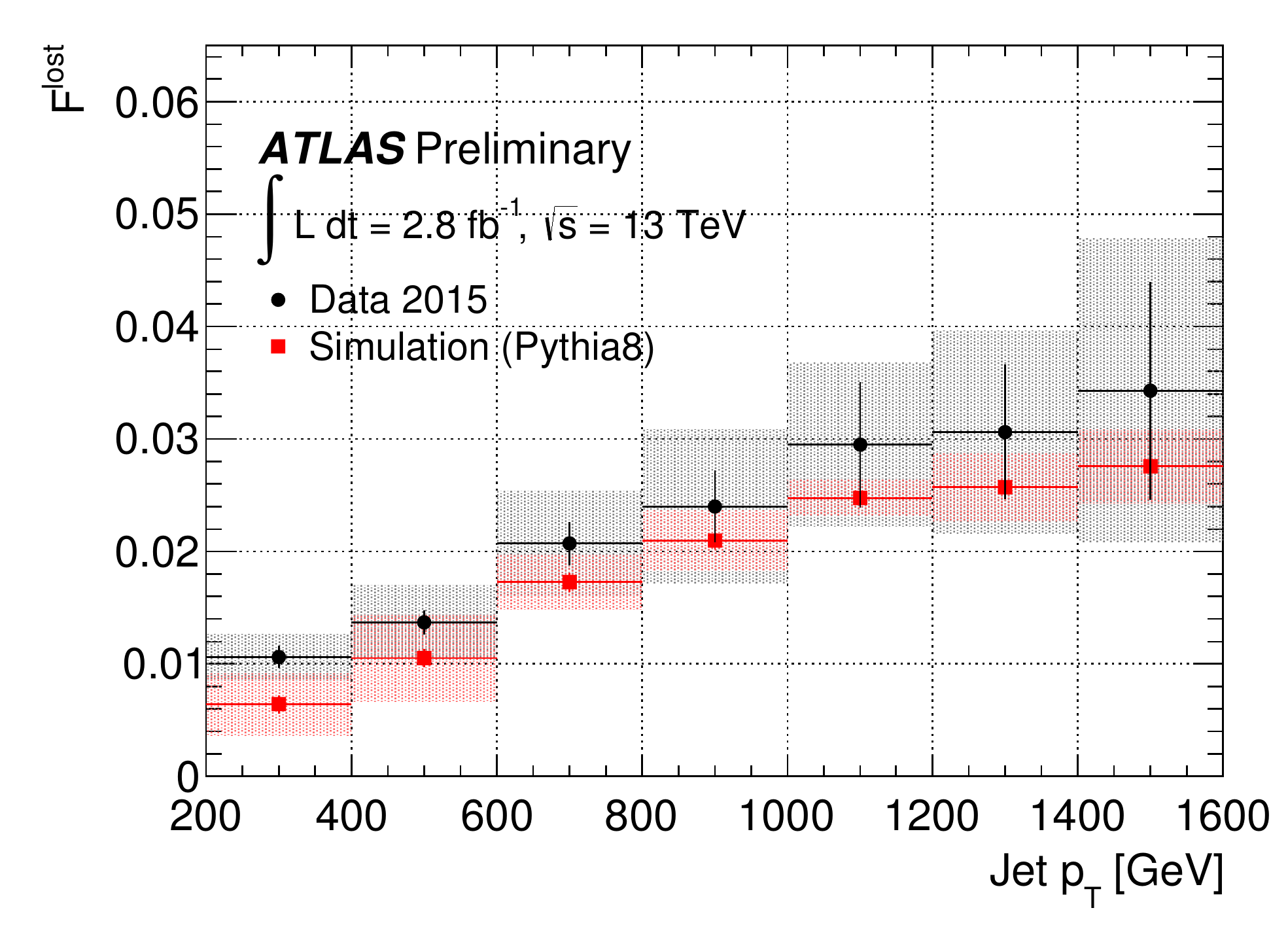}
\caption{\label{fig:floss} Fraction of lost tracks due to merged clusters, determined in data and simulation for varying values of $p_T^\mathrm{jet}$.  Shaded areas show total uncertainty including systematic uncertainties as described in section~\ref{sec:systs}.  From~\cite{pubnote}.}
\end{figure}

The fraction of lost tracks $F_\mathrm{loss}$ is given directly by the fit fraction of the multiple-track template.  The fit result for data can be seen in figure~\ref{fig:datafit} for two different $p_T$ bins.  Similar plots for Monte Carlo simulation are shown in~\cite{pubnote}.  The fraction of lost tracks depending on $p_T^\mathrm{jet}$ is shown in figure~\ref{fig:floss}, for data and simulation.  The loss fraction increases with $p_T^\mathrm{jet}$, and shows over the whole range agreement between data and simulation, within systematic uncertainties as outlined in section~\ref{sec:systs}.  The discrepancy between central values of data and simulation is approximately 25\%~\cite{pubnote}.

\section{Systematic Uncertainties}
\label{sec:systs}

The systematic uncertainties for simulation are dominated by Monte Carlo generator differences.  This uncertainty has been evaluated by comparing the fit results from \textsc{Pythia8}, \textsc{Sherpa} and \textsc{Herwig++} samples.  The relative systematic uncertainty ranges from 41\% at 200--400\;GeV to 5\% at 1000--1200\;GeV.  For details, see~\cite{pubnote}.

An additional uncertainty comes from the choice of fit region.  It was found that varying the upper edge of the region changes $F_\mathrm{lost}$, however only significantly in data.  A systematic uncertainty of the size of the maximal change in $F_\mathrm{lost}$ has been applied in each $p_T^\mathrm{jet}$ bin, varying between 12\% and 25\%.

In data, an uncertainty results from using low $p_T^\mathrm{jet}$ templates to fit high $p_T^\mathrm{jet}$ data.  A check has been carried out with a simulation of high statistics, and it was found that the fraction of clusters with three contributing tracks varies as a function of $p_T^\mathrm{jet}$.  This leads to a small bias in the resulting value of $F_\mathrm{lost}$ which has been taken into account as a systematic uncertainty.  The size of this uncertainty is between 11\% and 17\%.

% \section{Results \& Conclusions}

\section{Conclusions}
The tracking inefficiency in jet cores has been determined using measurements of energy deposition in the ATLAS pixel detector, on $\sqrt{s} = 13\;$TeV LHC data taken in 2015.  It was found that the fraction of lost tracks due to cluster merging is between $1\%$--$5\%$ for jet $p_T=200$--$1600\;$GeV.  The data and simulation are found to agree within 25\% in the investigated jet $p_T$ range.

\acknowledgments

The author would like to thank the conference organizers for an interesting and enjoyable conference, the authors of the presented results~\cite{pubnote} for their excellent work, and the ATLAS collaboration for the opportunity to present it.  The work is partly 
supported by the National Natural Science Foundation of China (Grant 
No. 11575200).

\end{document}